# Visualization and Control of Single Electron Charging in Bilayer Graphene Quantum Dots

Jairo Velasco Jr.[1,5,†,*], Juwon Lee[1,†], Dillon Wong[1†], Salman Kahn[1], Hsin-Zon Tsai[1], Joseph Costello[1], Torben Umeda[1], Takashi Taniguchi[4], Kenji Watanabe[4], Alex Zettl[1,2,3], Feng Wang[1,2,3] and Michael F. Crommie[1,2,3,*]

[1]*Department of Physics, University of California, Berkeley, California 94720, USA*
[2]*Materials Sciences Division, Lawrence Berkeley National Laboratory, Berkeley, California 94720, USA*
[3]*Kavli Energy NanoSciences Institute at the University of California, Berkeley and the Lawrence Berkeley National Laboratory, Berkeley, California 94720, USA*
[4]*National Institute for Materials Science, 1-1 Namiki, Tsukuba, 305-0044, Japan*
[5]*Department of Physics, University of California, Santa Cruz, California 95064, USA*
[†]*These authors contribute equally to this manuscript.*
*Email: jvelasc5@ucsc.edu & crommie@berkeley.edu*

# Abstract:

Graphene p-n junctions provide an ideal platform for investigating novel behavior at the boundary between electronics and optics that arise from massless Dirac fermions, such as whispering gallery modes and Veselago lensing. Bilayer graphene also hosts Dirac fermions, but they differ from single-layer graphene charge carriers because they are *massive,* can be gapped by an applied perpendicular electric field, and have very different pseudospin selection rules across a p-n junction. Novel phenomena predicted for these *massive* Dirac fermions at p-n junctions include anti-Klein tunneling, oscillatory Zener tunneling, and electron cloaked states. Despite these predictions there has been little experimental focus on the microscopic spatial behavior of massive Dirac fermions in the presence of p-n junctions. Here we report the experimental manipulation and characterization of massive Dirac fermions within bilayer graphene quantum dots defined by circular p-n junctions through the use of scanning tunneling microscopy-based (STM) methods. Our p-n junctions are created via a flexible technique that enables realization of exposed quantum dots in bilayer graphene/hBN heterostructures. These quantum dots exhibit sharp spectroscopic resonances that disperse in energy as a function of applied gate voltage. Spatial maps of these features show prominent concentric rings with diameters that can be tuned by an electrostatic gate. This behavior is explained by single-electron charging of localized states that arise from the quantum confinement of massive Dirac fermions within our exposed bilayer graphene quantum dots.

Recent advances in fabricating heterostructures of graphene and hexagonal boron nitride (hBN) have created new opportunities for fundamental study and control of Dirac fermions at the nanoscale. The ability to realize highly pristine, sharp interfaces and tunable potential landscapes in these systems has led to a renaissance in the study of electronic phenomena in graphene p-n junctions [1-4]. In particular, scanning tunneling microscopy (STM) of circular p-n junctions in graphene/hBN heterostructure devices has revealed exotic quasibound states known as whispering gallery modes [5-8]. Massive Dirac fermions in bilayer graphene (BLG) also possess attributes that make their behavior in p-n junction configurations equally novel, such as 100% barrier reflection and electron cloaking states [9-11]. Application of out-of-plane electric fields break inversion symmetry in BLG and can be used to create tunable-bandgap BLG p-n junction systems [12-15]. Transport experiments have been used to probe *massive* Dirac fermions in BLG p-n junctions [15-18], but studies that map their spatial behavior are lacking.

In order to locally explore the behavior of confined massive Dirac fermions we have utilized new STM-based techniques to create and characterize nanoscale circular p-n junctions in BLG [6, 19]. As depicted in Fig. 1a, our experimental setup consists of a Bernal-stacked BLG/hBN heterostructure placed on an $SiO_2$/Si wafer. The BLG is electrically grounded with a gold contact and its surface is exposed and accessible to an STM tip held at voltage $-V_S$ with respect to ground (the typical STM bias convention). A backgate voltage $V_G$ is applied to the heavily doped Si and can be used to tune the chemical potential ($\mu_g$) of the entire BLG flake. In order to create a local confinement potential in the graphene under these conditions, the STM tip is placed a few nanometers above the BLG and a $V_S = 5$ V voltage pulse is applied while holding the backgate at $\tilde{V}_G$ (see supplementary information for details). This process results in a buildup of space charge within the hBN below the location of the tip as shown in the inset of Fig. 1a [19]. This space charge (which acts as a local gate on the BLG) has a polarity opposite to the backgate voltage, is stable at T = 4K, and can be made either positive (which n-dopes the BLG) or negative (which p-dopes the BLG).

To verify that this procedure works for BLG/hBN we performed STM spectroscopy on BLG subjected to confinement potentials fabricated as described above (this technique is already well-established for single-layer graphene [6, 19]). Fig. 1b shows the differential conductance ($dI/dV_S$) of BLG measured with $V_G$ = 0 V after performing a tip-pulse procedure for various values of $\tilde{V}_G$. Each spectrum displays a ~130 mV gap-like feature at $V_S$ = 0 V that is known to arise from phonon-assisted inelastic tunneling [20, 21]. Adjacent to the phonon gap-like feature is a smaller dip that corresponds to the local charge neutrality point (CNP) of the BLG at the location of the tip pulse (marked by black arrows in Fig. 1b). When the CNP is to the right (left) of $V_S$ = 0 V, the BLG is p-doped (n-doped). A clear trend in the data of Fig. 1b is that a positive (negative) $\tilde{V}_G$ results in a shift of the CNP to the right (left) with the magnitude of the shift determined by the value of $\tilde{V}_G$. This behavior is consistent with previous experiments on single-layer graphene/hBN where positive (negative) backgate voltages p-doped (n-doped) graphene after application of tip-voltage pulses [7, 19]. The BLG behavior is also similar to single-layer graphene behavior in that the doping can be removed by applying a tip pulse while holding $\tilde{V}_G$ = 0 V, thus restoring the BLG to its pristine condition (green curve in Fig. 1b).

The modified electronic structure of locally gated BLG is best seen by measuring $dI/dV_S$ as a function of both $V_G$ and $V_S$ at the point where a tip-pulse has been applied. Fig. 2a shows BLG behavior at the site of a local gate created using $\tilde{V}_G$ = 60 V. One prominent feature in the data is a dark horizontal band around $V_S$ = 0 V that does not shift with $V_G$. This reflects the phonon-assisted inelastic tunneling feature present in the $dI/dV_S$ spectra shown in Fig. 1b. Another prominent feature is the set of bright (i.e., high $dI/dV_S$ signal) diagonal lines that shift in energy as a function of $V_G$. Four prominent lines with similar spacing appear near the middle of the plot and numerous fainter lines with smaller spacing and different slope appear to the left. The lines appear more clearly in Fig. 2b which shows the numerical derivative ($d^2I/dV_S^2$ ($V_G$, $V_S$)) of Fig. 2a. These features can also be seen in the "phonon gap" bias regime (-100 mV < $V_S$ < 100 mV) when the STM tunnel current is increased by a factor of two, revealing diamond-like features (see supplementary information for details). Fig. 2c shows a plot of $dI/dV_S$ spectra corresponding to line cuts at different values of $V_G$ marked by color-coded dashed lines in Fig. 2a. Sharp peaks appear in $dI/dV_S$ spectra for 46 V < $V_G$ < 52 V, but are absent when $V_G$ moves outside of this range. The energy spacing between $dI/dV_S$ peaks appears to increase with increasing $V_G$.

The spatial dependence of the electronic structure induced by locally gating BLG is obtained by performing $dI/dV_S$ maps in the regions subjected to a tip pulse. Figs. 3a and 3b show $dI/dV_S$ maps obtained at $V_G$ = 44V and 42V for a BLG region prepared similarly as the region discussed in Fig. 2. Multiple bright concentric rings can be seen in both $dI/dV_S$ maps. As the backgate voltage is lowered from $V_G$ = 44 V to $V_G$ = 42 V the number of rings and their spatial density increases. Figs. 3c, d show plots of the spatial and bias dependence of $d^2I/dV_S^2$ obtained along the dashed lines in Figs. 3a, b for $V_G$ = 44 V and 42V. Similar to Fig. 2b, $V_S$ = 0 V is bracketed by horizontal red and blue lines due to phonon-assisted inelastic tunneling. The most striking features in Figs. 3c, d are the multiple resonances that shift up together in energy as the distance from the origin is increased, thus forming nested parabola-like curves in $d^2I/dV_S^2$ ($V_S$, $x$). As $V_G$ is decreased from Fig. 3c to Fig. 3d, the number of resonances increases and the voltage spacing between them decreases.

Our experimental observations in Figs. 2 and 3 have some similarity to previous scanning gate microscopy (SGM) studies on quantum dots in carbon nanotubes [22, 23], InAs/InP nanowire heterostructures [24, 25], AlGaAs-GaAs heterostructures [26], and lithographically etched sheets of graphene [27]. In those experiments gate-tunable ring features were attributed to single-electron charging of a double-barrier quantum dot due to proximity of a conducting scanned probe tip. Our experimental observations can be similarly explained, but with notable differences regarding the mechanism for charge confinement. Massive Dirac fermions in circular BLG p-n junctions are confined via two mechanisms: (1) novel barrier reflection at the p-n junction (anti-Klein tunneling) [9, 28]; and (2) suppression of states via an electric-field-tunable band gap [12]. Confinement mechanism (1) causes normally-incident electrons to reflect and become trapped within a circular BLG p-n junction, contributing to sharp quantum dot (QD) eigenstates [28]. Confinement mechanism (2) causes a tunnel barrier to emerge at the p-n junction due to an out-of-plane electric field that induces a local bandgap [29-31] (see supplementary information for details), thus resulting in confinement.

Confinement mechanism (2) is most likely the dominant mechanism in our experimental system. This is supported by the fact that the doping level achieved by our *local* gating technique (40 to 100 meV) is comparable to doping levels achieved previously in BLG systems by *global* backgating that resulted in BLG energy gaps [32, 33]. Prior theoretical work and transport measurements indicate that in this regime anti-Klein tunneling is suppressed because the pseudospin selection rules across the p-n junction are modified by the presence of the energy gap [16, 34, 35]. The gate-dependent electronic structure of the QD can be understood schematically from the drawings in Figs. 2d-g. Fig. 2d corresponds to the highest gate voltage ($V_G$ = 54 V) which is represented by the dashed purple line in Fig. 2a and results in the top $dI/dV_s$ spectrum in Fig. 2c.

Here the high gate voltage causes the BLG $\mu_g$ to lie above the dome-like potential (i.e., the spatially-dependent CNP) induced by trapped negative charges in the hBN layer. The quantum dot states lie far below $\mu_g$ and so are not seen in the $dI/dV_S$ spectrum. As the gate voltage is reduced to $V_G = 50$ V (dashed green line in Fig. 2a) $\mu_g$ lowers and intersects the top of the dome potential (Fig. 2e). Because the states here lie in the narrowest region of the dome they have larger energetic separation, as seen by the peak spacing in the corresponding $dI/dV_S$ spectrum of Fig. 2c (green curve). When the gate voltage is reduced to $V_G = 46$ V, $\mu_g$ lowers into a wider section of the dome potential (i.e., that has a wider p-n junction boundary) that corresponds to states with smaller energy separation (Fig. 2f). This is reflected in the more closely spaced peaks of the corresponding $dI/dV_S$ spectrum in Fig. 2c (orange curve). As the gate voltage is reduced further ($V_G = 44$ V), $\mu_g$ lowers and the quantum dot grows even wider (Fig. 2g). The quantum dot levels eventually become so close in energy that they cannot be separately detected, as seen in the corresponding $dI/dV_S$ spectrum of Fig. 2c (red curve). The background signal observed in Fig. 2c may be partially attributed to direct tunneling from the tip to the bulk BLG outside of the potential well.

The distinctive upward slope (as $V_G$ is increased) of the resonant lines in Figs. 2a, b arises due to the single-electron charging nature of these peaks. Similar features have been observed in single-layer graphene quantum dots in zero and finite perpendicular magnetic field [5, 36], but there are differences for BLG. In BLG dots the quantum level detection mechanism is equivalent to charging a double-barrier quantum dot. The first barrier is the vacuum gap between the STM tip and the dot while the second barrier arises from the BLG energy gap at the p-n junction between the interior (QD) and exterior (bulk) BLG (this barrier does not exist for single-layer graphene whispering gallery mode systems). For $V_S > 0$ V the STM tip acts both as a charge reservoir for the double barrier quantum dot as well as the initiator of single-electron charging events where charge is transferred from the dot to the exterior BLG. For $V_S < 0$ V the QD receives an electron from the BLG bulk, thus resulting in the distinctive dependence of the charging peaks on $V_G$ and $V_S$ as shown in Figs. 2a, b.

The details of this process are sketched in Fig. 4 where the unoccupied QD eigenstates (dashed lines) are separated from occupied QD eigenstates (solid lines) by an electron addition energy $E_{add}$[37]. When the lateral distance between the STM tip and the QD center is decreased from $R_1$ to $R_2$, as shown in Figs. 4a-b, the QD energy levels are raised due to increased tip-gating efficiency (i.e., greater capacitive coupling between the QD and the tip, $C_{tip}(r)$, as represented in Fig. 4e) [38]. Fig. 4c shows the energy level diagram as the STM bias voltage is increased slightly from $V_S$ to $V_S + \delta V_S$, which further lifts the QD energy levels and allows an electron to tunnel out of the QD and into the BLG bulk, thus changing the net charge on the QD. Since the rate of tunneling between the tip and QD is determined by the number of empty QD states available within the bias window, the resulting enhancement in empty states produces a jump in current and a sharp peak in the $dI/dV_S$ spectrum (as shown in Fig. 4d) each time an occupied QD level crosses $\mu_g$ [37, 39-42]. When QD levels are very closely spaced and a large asymmetry exists between the tip/QD and QD/bulk-graphene tunnel barriers then a corresponding asymmetry can exist between the charging and *direct* tunneling signals that occur as the tip chemical potential crosses a QD state [42-45].

These Coulomb blockade resonances provide information regarding the electronic structure of BLG QDs that confine massive Dirac fermions. For example, the voltage spacing between charging peaks in $dI/dV_S$ spectra are related to $E_{add}$, where $E_{add} = \Delta E + E_C$, $\Delta E$ is the single-particle orbital splitting set by the confinement potential, and $E_C = e^2/C$ is the charging energy required to add an electron to a quantum dot of total capacitance $C$ [37]. Both $\Delta E$ and $E_C$ are expected to decrease when the QD diameter is increased, consistent with the trend observed in Fig.

2 where the voltage spacing between Coulomb peaks in d$I$/d$V_S$ spectra decreases as the effective size of the quantum dot increases. d$I$/d$V_S$ spectra in Fig. 3c show equal voltage spacing for all resonances except between the fourth and the fifth, where the voltage spacing is largest. This is consistent with the expected four-fold degeneracy in QD energy levels due to spin and valley symmetries which allow $\Delta E$ to make a non-zero contribution only after all orbitally degenerate levels have been filled. The Coulomb peaks appear more equally spaced in the larger quantum dot of Fig. 3d, signifying a transition to a regime where $E_C$ is the dominant contribution to $E_{add}$.

This interpretation of the observed d$I$/d$V_S$ peaks is further validated by estimating the level spacing $\Delta E$ for confined massive Dirac fermions and comparing it to an estimated $E_{add}$. Here we approximate the confinement potential with a parabola having level spacing $\hbar\omega$, where $\omega = \sqrt{\frac{k}{m^*}}$, $m^* = 0.03 m_e$ (the BLG electron effective mass) [46], and $k$ is the curvature of the electrostatic potential. For $k$ we use the experimental value extracted from our previous study on tip-induced QDs in single-layer graphene [6]. Because of stronger screening in BLG compared to single-layer graphene [12], this $k$ value represents an upper limit. The actual quantum dot level spacing is thus expected to be less than the calculated level spacing of $\Delta E \sim 5$ meV, which is significantly smaller than the estimated BLG QD charging energy of $E_C \sim 45$ meV (estimated using a 2D disc capacitance model [39]). Analysis of the Coulomb-diamond-like features observed at small $V_s$ (Fig. S1b) yield an $E_{add}$ of ~27meV, consistent with our estimate for $E_C$ and $\Delta E$ (see supplementary information for details). These low-bias spectra are expected to exhibit the electric field induced gap in addition to Coulomb diamonds, thus complicating the assignment of the diamond-like features directly to $E_{add}$. Nonetheless, charging behavior is seen to dominate the response of the BLG QD ($E_C \gg \Delta E$), unlike the case for single-layer QDs where the situation is reversed. This striking difference between single-layer graphene and BLG quantum dots could be attributed to the larger level spacing for single layer graphene QDs (~15 meV [6]) as well as the larger resistance expected from the barrier surrounding each BLG QD.

In conclusion, we have taken the first steps towards visualizing and manipulating massive Dirac fermion states confined within circular p-n junctions. We observe sharp resonances and prominent concentric rings with gate-tunable diameters that can be explained by charging of single-electron states that arise from confinement of massive Dirac fermions within a circular p-n junction. The Coulomb charging energy dominates over the orbital-level spacing energy in the BLG QDs synthesized up to now. We expect that thinning the hBN insulator thickness should allow steeper electrostatic potentials, thus enabling smaller BLG quantum dots with orbital-level spacings that surpass the charging energy. Such dots should facilitate more direct imaging of quantum interference effects in QD wave functions, thus providing a deeper understanding of the physics of massive Dirac fermions in BLG.

**Acknowledgments:**

This research was supported by the sp$^2$ program (KC2207) (STM measurement and instrumentation) funded by the Director, Office of Science, Office of Basic Energy Sciences Materials Sciences and Engineering Division, of the US Department of Energy under Contract No. DE-AC02-05CH11231. For the graphene characterization we used the Molecular Foundry at LBNL, which is funded by the Director, Office of Science, Office of Basic Energy Sciences, Scientific User Facilities Division, of the US Department of Energy under Contract No. DE-AC02-05CH11231. Support was also provided by National Science Foundation award DMR-1206512 (device fabrication, image analysis). D.W. was supported by the Department of Defense (DoD) through the National Defense Science & Engineering Graduate Fellowship (NDSEG) Program, 32 CFR 168a.

**Author Contributions:**

J.V.J., J.L., and D.W. conceived the work and designed the research strategy. J.V.J. and J.L. performed data analysis. S.K., J.C., T.U., and A.Z. facilitated sample fabrication. J.L., J.V.J, and D.W. carried out STM/STS measurements. K.W. and T.T. synthesized the h-BN samples. M.F.C. supervised the STM/STS experiments. J.L., J.V.J., D.W, and M.F.C formulated the theoretical model. J.V.J., J.L., D.W., and M.F.C. co-wrote the manuscript. J.V.J. and M.F.C. coordinated the collaboration. All authors discussed the results and commented on the paper.

**Supporting Information:**

(1) Description of tip pulsing procedure, (2) Analysis of low-bias diamond-like features, and (3) Cross section schematic of bilayer graphene p-n junction.

**Sample Fabrication:** Our samples were fabricated using a transfer technique developed by Zomer *et al.* [47] that uses 60-100 nm thick hBN crystals (synthesized by Taniguchi and Watanabe) and 300 nm thick $SiO_2$ as the dielectric for electrostatic gating. Bilayer graphene was mechanically exfoliated from graphite and deposited onto methyl methacrylate (MMA) before being transferred onto hBN previously exfoliated onto a heavily doped $SiO_2$/Si wafer. The bilayer graphene was electrically grounded through a Ti (10 nm)/Au (40-100 nm) electrode deposited via electron-beam evaporation using a shadow mask. Devices were annealed in $Ar/H_2$ gas at 350°C before being transferred into our Omicron ultra-high vacuum (UHV) low-temperature STM. A second anneal was performed overnight at 250-400°C and $10^{-11}$ torr.

**Scanning tunneling microscopy and spectroscopy measurements:** All STM measurements were performed at T = 4.8 K with platinum iridium STM tips calibrated against the Au(111) Shockley surface state. STM topographic and $dI/dV_S$ images were obtained at constant current with sample bias $V_S$ defined as the negative of the voltage applied to the STM tip with respect to the grounded bilayer graphene sample. Scanning tunneling spectroscopy (STS) measurements were performed by lock-in detection of the a.c. tunnel current induced by a modulated voltage (1-6 mV at 613.7 Hz) added to $V_S$. $d^2I/dV_S^2$ was obtained through numerical differentiation. The results presented here were reproduced with numerous STM tips on more than 15 circular p-n junctions.

**Figure 1. Local doping of bilayer graphene (BLG).** (a) Schematic for using a scanning tunneling microscope (STM) to locally dope a Bernal-stacked BLG/hexagonal boron nitride (hBN) heterostructure resting upon a SiO$_2$/Si wafer. A voltage pulse is applied to an STM tip positioned a few nanometers above the graphene surface while the Si backgate is held at a voltage $\tilde{V}_G$. This ionizes defects within the hBN (inset), resulting in a buildup of space charge that acts as a local embedded gate on the BLG. (b) d$I$/d$V_S$ spectra showing shifts in the BLG charge neutrality point (CNP) due to local doping. Each d$I$/d$V_S$ curve was obtained with the backgate held at $V_G = 0$ V. Initial tunneling parameters: $V_s = 0.5$ V, $I = 0.5$ nA, $V_{ac} = 5$ mV.

**Figure 2. Scanning tunneling spectroscopy (STS) of locally doped bilayer graphene.** (a) d$I$/d$V_S$($V_G,V_S$) plot consisting of d$I$/d$V_S$ spectra measured at the center of a locally p-doped region of the BLG. The bright diagonal lines are peaks in d$I$/d$V_S$ spectra that shift in energy as a function of $V_G$. (b) d$^2I$/d$V_S^2$($V_G,V_S$) produced by numerically differentiating (a). (c) d$I$/d$V_S$ spectra corresponding to line cuts at $V_G$ indicated by dashed lines in (a). Each spectrum is offset for clarity by 0.2 a.u from the spectrum below it (initial tunneling parameters for all d$I$/d$V_S$ spectra: $V_S = 0.5$ V, $I = 0.5$ nA, $V_{ac} = 5$ mV). (d-g) Schematic illustration of how the backgate can be used to create and tune a circular p-n junction in BLG. (g) shows a region of the BLG that is locally p-doped by an STM tip voltage pulse. Increasing $V_G$ gradually n-dopes the BLG and reduces the area that is p-doped, as shown in (f) and (e). For sufficiently large values of $V_G$, the BLG will become completely n-doped, as shown in (d).

**Figure 3. Spatially resolved STS of circular p-n junctions in bilayer graphene.** (a, b) d$I$/d$V_S$ maps of locally p-doped bilayer graphene at two different backgate voltages. (c, d) d$^2I$/d$V_S^2$($x,V_S$)

measured along dashed lines in **a**, **b**. Horizontal dashed lines in **c**, **d** indicate $V_S$ for measurements shown in **a**, **b**. Arrow in **c** highlights increased spacing between fourth and fifth levels. Initial tunneling parameters for d$I$/d$V_S$ spectra: $V_S$= 0.5 V, $I$=0.5n, $V_{ac}$=5mV. (e, f) sketch shows spatially varying charge neutrality point ($E_{CNP}$) in locally p-doped bilayer graphene. $E_{CNP}$ is raised locally relative to the chemical potential of graphene, $\mu_g$. Lowering $V_G$ raises $E_{CNP}$ relative to $\mu_g$, expanding the area of BLG that is p-doped.

**Figure 4. Single-electron charging of a double-barrier quantum dot in BLG.** (a-c) Schematic illustration of how the STM tip can be used to control and detect the charge state of a BLG quantum dot (QD). From (a) to (b) the STM tip moves toward the QD interior ($R_2 < R_1$) while the sample bias $V_S$ remains constant. From (b) to (c) the lateral distance is held constant but a slightly higher sample bias $V_S + \delta V_S$ is applied. Energetic alignments between the STM tip, the QD, and the BLG at tip positions and sample biases are shown. $E_{add}$ is the electron addition energy gap between unoccupied levels (dashed lines) and occupied levels (solid lines). In **a**, **b** the gating efficiency between the tip and QD increases as the tip moves toward the QD interior which lifts the QD levels. In (c) the sample bias $V_S$ is increased which further shifts the QD potential and ionizes the QD. This results in a sudden increase in the number of empty QD states in the bias window available for tunneling, causing a jump in (d) the tunneling conductance and producing a peak in d$I$/d$V_S$. (e) An effective circuit diagram of the system when the STM tip is positioned above the QD. The tip (bulk BLG) is coupled to the QD through capacitance $C_{tip}$ ($C_{bulk}$) and tunneling resistance $R_{tip}$ ($R_{bulk}$). The Si backgate is coupled to the QD ($C_{QD}$) as well as the bulk bilayer graphene ($C_{bulk}$).

# Figure 1

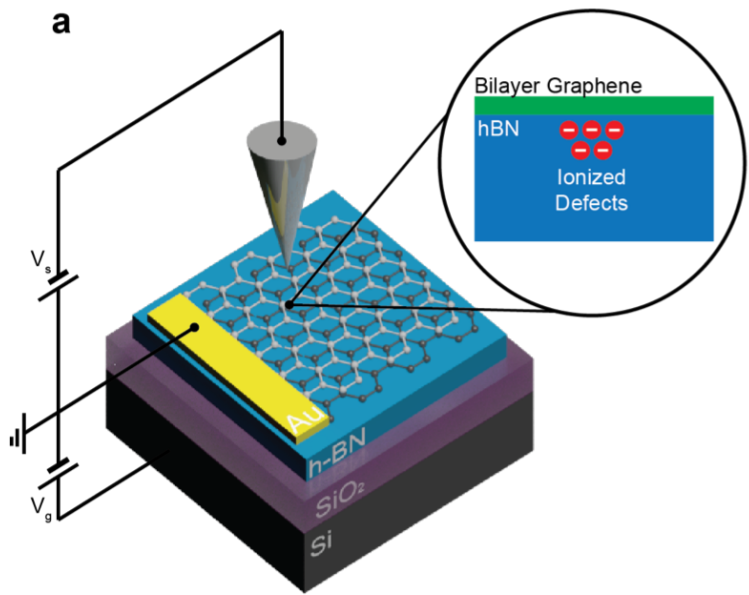
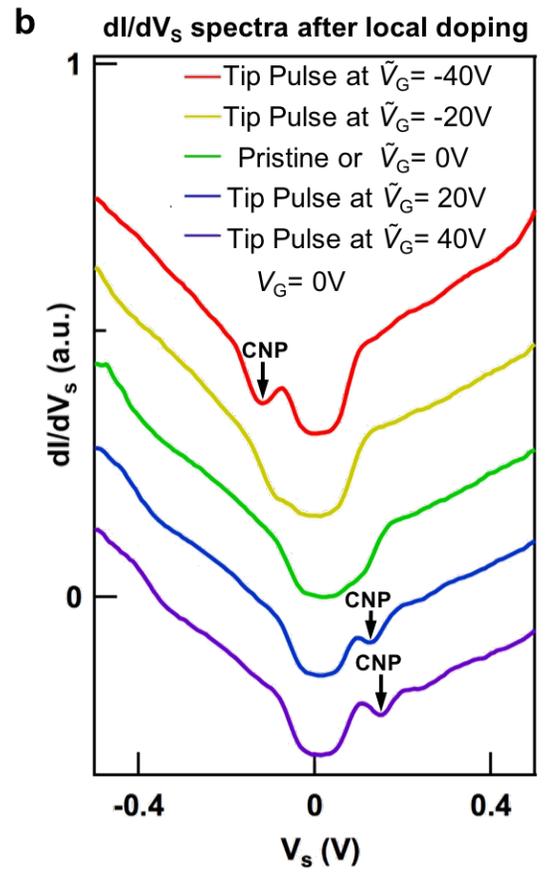

**Figure 2**

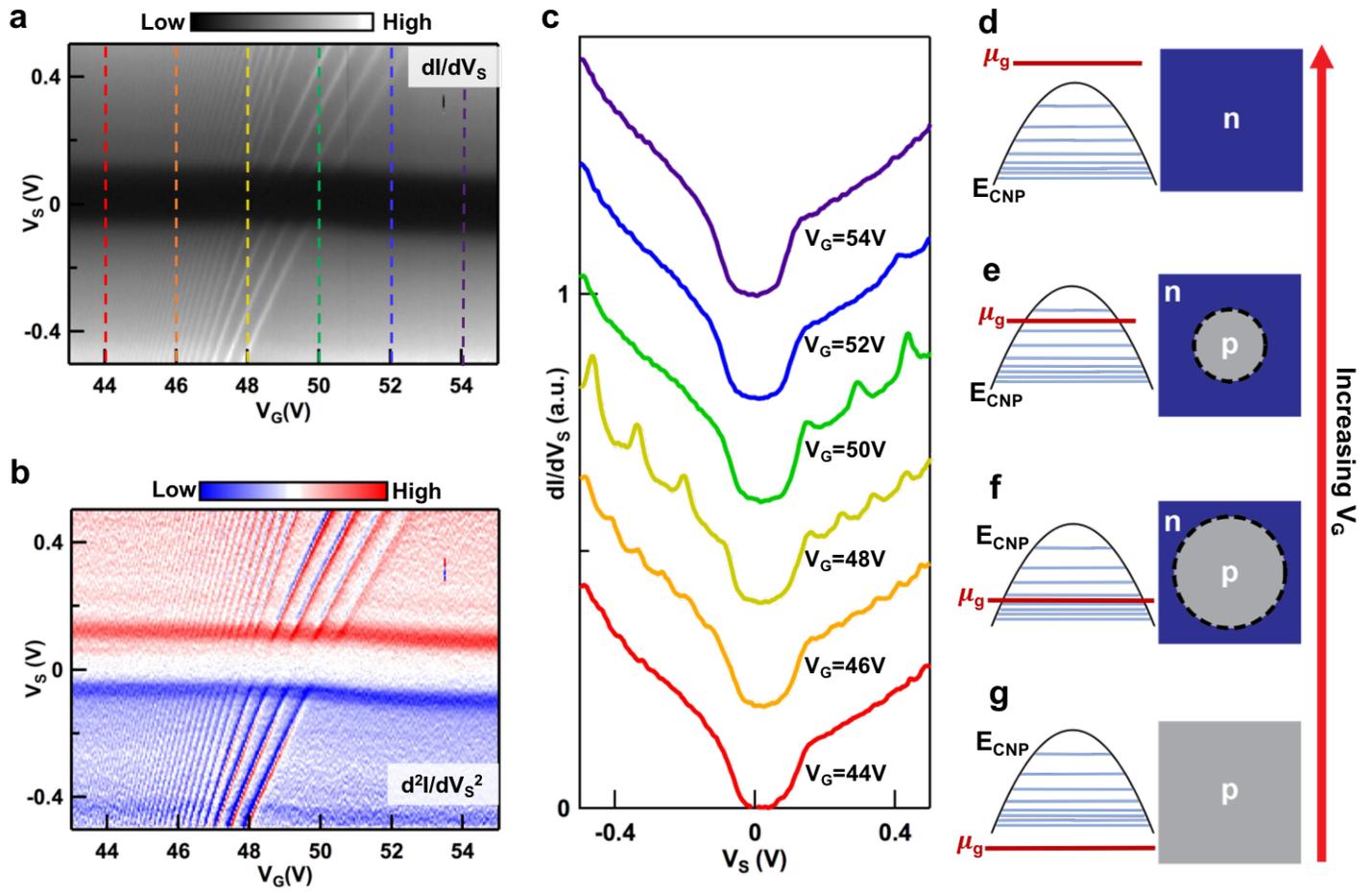

# Figure 3

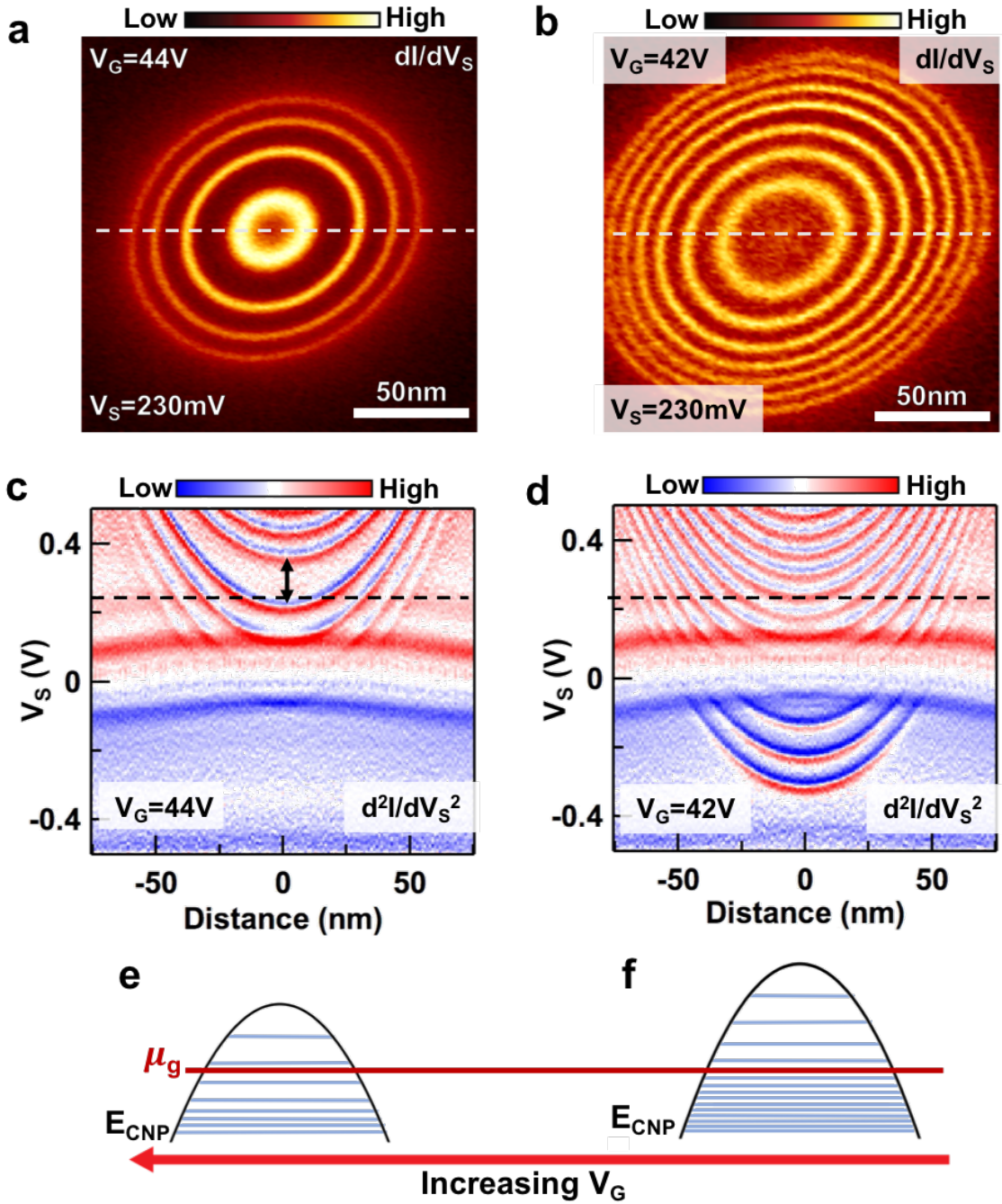

**Figure 4**

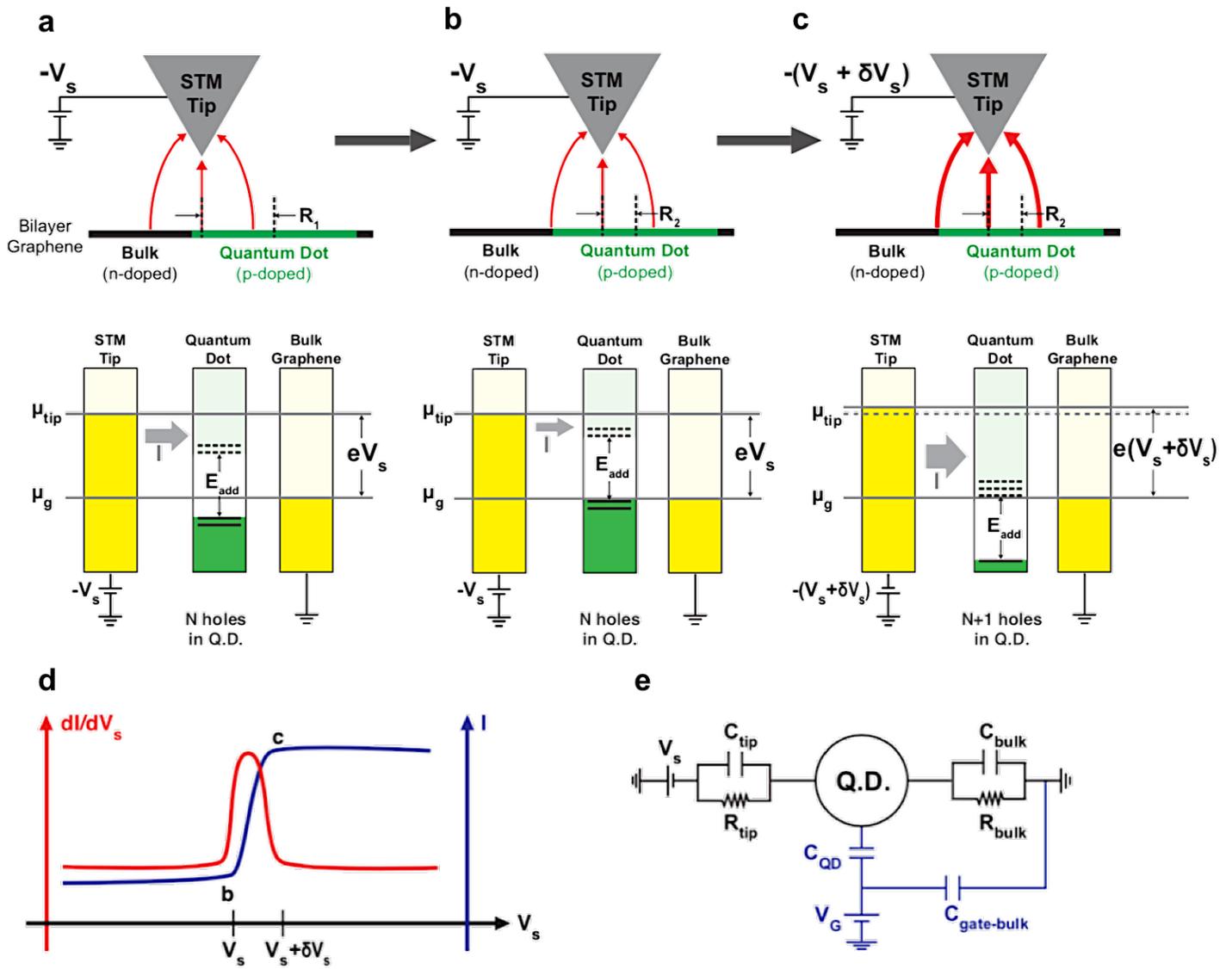

**References**

(1) Lee, G.-H.; Park, G.-H.; Lee, H.-J. *Nat. Phys.* **2015,** 11, 925.

(2) Chen, S.; Han, Z.; Elahi, M. M.; Habib, K. M. M.; Wang, L.; Wen, B.; Gao, Y.; Taniguchi, T.; Watanabe, K.; Hone, J.; Ghosh, A. W.; Dean, C. R. *Science* **2016,** 353, 1522.

(3) Liu, M.-H.; Gorini, C.; Richter, K. *Phys. Rev. Lett.* **2017,** 118, 066801.

(4) Rickhaus, P.; Makk, P.; Liu, M.-H.; Tóvári, E.; Weiss, M.; Maurand, R.; Richter, K.; Schönenberger, C. *Nat. Commun.* **2015,** 6, 6470.

(5) Zhao, Y.; Wyrick, J.; Natterer, F. D.; Rodriguez-Nieva, J. F.; Lewandowski, C.; Watanabe, K.; Taniguchi, T.; Levitov, L. S.; Zhitenev, N. B.; Stroscio, J. A. *Science* **2015,** 348, 672.

(6) Lee, J.; Wong, D.; Velasco Jr, J.; Rodriguez-Nieva, J. F.; Kahn, S.; Tsai, H.-Z.; Taniguchi, T.; Watanabe, K.; Zettl, A.; Wang, F.; Levitov, L. S.; Crommie, M. F. *Nat. Phys.* **2016,** 12, 1032.

(7) Ghahari, F.; Walkup, D.; Gutiérrez, C.; Rodriguez-Nieva, J. F.; Zhao, Y.; Wyrick, J.; Natterer, F. D.; Cullen, W. G.; Watanabe, K.; Taniguchi, T.; Levitov, L. S.; Zhitenev, N. B.; Stroscio, J. A. *Science* **2017,** 356, 845.

(8) Jiang, Y.; Mao, J.; Moldovan, D.; Masir, M. R.; Li, G.; Watanabe, K.; Taniguchi, T.; Peeters, F. M.; Andrei, E. Y. *Nat. Nanotechnol.* **2017,** 12, 1045.

(9) Katsnelson, M. I.; Novoselov, K. S.; Geim, A. K. *Nat. Phys.* **2006,** 2, 620.

(10) Shytov, A. *arXiv:1506.02839* **2015**.

(11) Gu, N.; Rudner, M.; Levitov, L. *Phys. Rev. Lett.* **2011,** 107, 156603.

(12) McCann, E. *Phys. Rev. B* **2006,** 74, 161403.

(13) Oostinga, J. B.; Heersche, H. B.; Liu, X. L.; Morpurgo, A. F.; Vandersypen, L. M. K. *Nat. Mater.* **2008,** 7, 151.

(14) Zhang, Y. B.; Tang, T. T.; Girit, C.; Hao, Z.; Martin, M. C.; Zettl, A.; Crommie, M. F.; Shen, Y. R.; Wang, F. *Nature* **2009,** 459, 820.

(15) Jing, L.; Velasco Jr, J.; Kratz, P.; Liu, G.; Bao, W.; Bockrath, M.; Lau, C. N. *Nano Lett.* **2010,** 10, 4000.

(16) Varlet, A.; Liu, M.-H.; Krueckl, V.; Bischoff, D.; Simonet, P.; Watanabe, K.; Taniguchi, T.; Richter, K.; Ensslin, K.; Ihn, T. *Phys. Rev. Lett.* **2014,** 113, 116601.

(17) Lee, K.; Lee, S.; Eo, Y. S.; Kurdak, C.; Zhong, Z. *Phys. Rev. B* **2016,** 94, 205418.

(18) Renjun Du, M.-H. L., Jens Mohrmann, Fan Wu, Ralph Krupke, Hilbert v. Lohneysen, Klaus Richter, Romain Danneau. *arXiv:1703.07260v2,* **2018**.

(19) Velasco, J.; Ju, L.; Wong, D.; Kahn, S.; Lee, J.; Tsai, H.-Z.; Germany, C.; Wickenburg, S.; Lu, J.; Taniguchi, T.; Watanabe, K.; Zettl, A.; Wang, F.; Crommie, M. F. *Nano Lett.* **2016,** 16, 1620.

(20) Zhang, Y.; Brar, V. W.; Wang, F.; Girit, C.; Yayon, Y.; Panlasigui, M.; Zettl, A.; Crommie, M. F. *Nat. Phys.* **2008,** 4, 627.

(21) Decker, R. g.; Wang, Y.; Brar, V. W.; Regan, W.; Tsai, H.-Z.; Wu, Q.; Gannett, W.; Zettl, A.; Crommie, M. F. *Nano Lett.* **2011,** 11, 2291.

(22) Bockrath, M.; Liang, W.; Bozovic, D.; Hafner, J. H.; Lieber, C. M.; Tinkham, M.; Park, H. *Science* **2001,** 291, 283.

(23) Woodside, M. T.; McEuen, P. L. *Science* **2002,** 296, 1098.

(24) Bleszynski-Jayich, A. C.; Fröberg, L. E.; Björk, M. T.; Trodahl, H. J.; Samuelson, L.; Westervelt, R. M. *Phys. Rev. B* **2008,** 77, 245327.

(25) Bleszynski, A. C.; Zwanenburg, F. A.; Westervelt, R. M.; Roest, A. L.; Bakkers, E. P. A. M.; Kouwenhoven, L. P. *Nano Lett.* **2007,** 7, 2559.

(26) Pioda, A.; Kičin, S.; Ihn, T.; Sigrist, M.; Fuhrer, A.; Ensslin, K.; Weichselbaum, A.; Ulloa, S. E.; Reinwald, M.; Wegscheider, W. *Phys. Rev. Lett.* **2004,** 93, 216801.

(27) Schnez, S.; Güttinger, J.; Huefner, M.; Stampfer, C.; Ensslin, K.; Ihn, T. *Phys. Rev. B* **2010,** 82, 165445.

(28) Matulis, A.; Peeters, F. M. *Phys. Rev. B* **2008,** 77, 115423.

(29) Pereira, J. M.; Vasilopoulos, P.; Peeters, F. M. *Nano Lett.* **2007,** 7, 946.

(30) Allen, M. T.; Martin, J.; Yacoby, A. *Nat. Commun.* **2012,** 3, 934.

(31) Marius Eich, F. H., Riccardo Pisoni, Hiske Overweg, Yongjin Lee, Peter Rickhaus, Kenji Watanabe, Takashi Taniguchi, Manfred Sigrist, Thomas Ihn, Klaus Ensslin *arXiv:1803.02923* **2018**.

(32) Rutter, G. M.; Jung, S.; Klimov, N. N.; Newell, D. B.; Zhitenev, N. B.; Stroscio, J. A. *Nat. Phys.* **2011,** 7, 649.

(33) Yankowitz, M.; Wang, J. I. J.; Li, S.; Birdwell, A. G.; Chen, Y.-A.; Watanabe, K.; Taniguchi, T.; Quek, S. Y.; Jarillo-Herrero, P.; LeRoy, B. J. *APL Mater.* **2014,** 2, 092503.

(34) Varlet, A.; Liu, M.-H.; Bischoff, D.; Simonet, P.; Taniguchi, T.; Watanabe, K.; Richter, K.; Ihn, T.; Ensslin, K. *Phys. Status Solidi (RRL)* **2016,** 10, 46.

(35) Park, S.; Sim, H. S. *Phys. Rev. B* **2011,** 84, 235432.

(36) Freitag, N. M.; Chizhova, L. A.; Nemes-Incze, P.; Woods, C. R.; Gorbachev, R. V.; Cao, Y.; Geim, A. K.; Novoselov, K. S.; Burgdörfer, J.; Libisch, F.; Morgenstern, M. *Nano Lett.* **2016,** 16, 5798.

(37) Sohn, L. L.; Kouwenhoven, L. P.; Schön, G., *Mesoscopic Electron Transport*. Springer Netherlands: 1997.

(38) Brar, V. W.; Decker, R.; Solowan, H.-M.; Wang, Y.; Maserati, L.; Chan, K. T.; Lee, H.; Girit, C. O.; Zettl, A.; Louie, S. G.; Cohen, M. L.; Crommie, M. F. *Nat. Phys.* **2011,** 7, 43.

(39) Kouwenhoven, L. P.; Marcus, C. M.; McEuen, P. L.; Tarucha, S.; Westervelt, R. M.; Wingreen, S., *Electron Transport in Quantum Dots*. Springer: Kluwer, 1997.

(40) Kastner, M. A. *Rev. of Mod. Phys.* **1992,** 64, 849.

(41) Averin, D. V.; Korotkov, A. N.; Likharev, K. K. *Phys. Rev. B* **1991,** 44, 6199.

(42) LeRoy, B. J.; Kong, J.; Pahilwani, V. K.; Dekker, C.; Lemay, S. G. *Phys. Rev. B* **2005,** 72, 075413.

(43) Park, J. Electron Transport in Single Molecule Transistor. UNIVERSITY OF CALIFORNIA, BERKELEY, 2003.

(44) Bakkers, E. P. A. M.; Hens, Z.; Zunger, A.; Franceschetti, A.; Kouwenhoven, L. P.; Gurevich, L.; Vanmaekelbergh, D. *Nano Lett.* **2001,** 1, 551.

(45) Jdira, L.; Liljeroth, P.; Stoffels, E.; Vanmaekelbergh, D.; Speller, S. *Phys. Rev. B* **2006,** 73, 115305.

(46) Henriksen, E. A.; Jiang, Z.; Tung, L. C.; Schwartz, M. E.; Takita, M.; Wang, Y. J.; Kim, P.; Stormer, H. L. *Phys. Rev. Lett.* **2008,** 100, 087403.

(47) Zomer, P. J.; Dash, S. P.; Tombros, N.; van Wees, B. J. *Appl. Phys. Lett.* **2011,** 99, 232104.

**For TOC only**

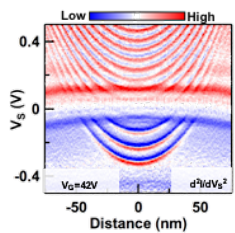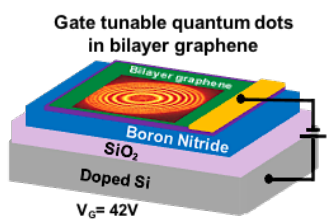